\documentclass[graybox]{svmult}

\usepackage{mathptmx}       % selects Times Roman as basic font
\usepackage{helvet}         % selects Helvetica as sans-serif font
\usepackage{courier}        % selects Courier as typewriter font
\usepackage{type1cm}        % activate if the above 3 fonts are
\usepackage{makeidx}         % allows index generation
\usepackage{graphicx}        % standard LaTeX graphics tool
\usepackage{multicol}        % used for the two-column index
\usepackage[bottom]{footmisc}% places footnotes at page bottom

\usepackage{amsmath}
\usepackage{amssymb}
\newcommand{\Tt}[1]{\mathbf{#1}}

\makeindex             % used for the subject index
\begin{document}

\title*{Parallel-in-Space-and-Time Simulation of the Three-Dimensional, Unsteady Navier-Stokes Equations for 
Incompressible Flow}
\titlerunning{Space-Time-Parallel Navier-Stokes} 

\author{Roberto Croce and Daniel Ruprecht and Rolf Krause}
\authorrunning{R. Croce, D. Ruprecht, R. Krause} 
\institute{Roberto Croce \at Institute of Computational Science, Via Giuseppe Buffi 13, CH-6906 Lugano, Switzerland,
\email{roberto.croce@usi.ch}
\and Daniel Ruprecht \at Institute of Computational Science, Via Giuseppe Buffi 13, CH-6906 Lugano, Switzerland,
\email{daniel.ruprecht@usi.ch}
\and Rolf Krause \at Institute of Computational Science, Via Giuseppe Buffi 13, CH-6906 Lugano, Switzerland,
\email{rolf.krause@usi.ch} }
\maketitle

%
% ---> 10 pages maximum.
%

\abstract*{% Abstract
In this paper we combine the Parareal parallel-in-time method together with spatial parallelization
and investigate this space-time parallel scheme by means of solving the three-dimensional incompressible
Navier-Stokes equations. Parallelization of time stepping provides a new direction of parallelization
and allows to employ additional cores to further speed up simulations after spatial parallelization
has saturated. We report on numerical experiments performed on a Cray XE6, simulating a driven
cavity flow with and without obstacles. Distributed memory parallelization is used in both space and time,
featuring up to 2,048 cores in total. It is confirmed that the space-time-parallel method can provide speedup 
beyond the saturation of the spatial parallelization.
}

\abstract{% Abstract
In this paper we combine the Parareal parallel-in-time method together with spatial parallelization
and investigate this space-time parallel scheme by means of solving the three-dimensional incompressible
Navier-Stokes equations. Parallelization of time stepping provides a new direction of parallelization
and allows to employ additional cores to further speed up simulations after spatial parallelization
has saturated. We report on numerical experiments performed on a Cray XE6, simulating a driven
cavity flow with and without obstacles. Distributed memory parallelization is used in both space and time,
featuring up to 2,048 cores in total. It is confirmed that the space-time-parallel method can provide speedup 
beyond the saturation of the spatial parallelization.
}

\section{Introduction}
Simulating three-dimensional flows by numerically solving the
time-dependent Navier-Stokes equations leads to huge
computational costs. In order to obtain a reasonable time-to-solution,
massively parallel computer systems have to be utilized. This requires
sufficient parallelism to be identifiable in the employed solution
algorithms. Decomposition of the spatial computational domain is by
now a standard technique and has proven to be extremely
powerful. Nevertheless, for a fixed problem size, this approach can
only push the time-to-solution down to some fixed threshold, below
which the computation time for each subdomain becomes comparable to
the communication time. While pure spatial parallelization can provide
satisfactory runtime reduction, time-critical applications may require
larger speedup and hence need additional directions of parallelism in
the used numerical schemes.

One approach that has received increasing attention over recent years
is parallelizing the time-stepping procedure typically used to solve
time-dependent problems. A popular algorithm for this is
\emph{Parareal}, introduced in~\cite{LionsEtAl2001} and
comprehensively analyzed in~\cite{Gander2007}. Its performance has
been investigated for a wide range of problems, see for example the
references in~\cite{Minion2010,RuprechtKrause2012}. A first
application to the 2D-Navier-Stokes equations, focussing on stability
and accuracy without reporting runtimes, can be found
in~\cite{FischerEtAl2005}. Some experiments with a combined
Parareal/domain-decomposition parallelization  for the two-dimensional
Navier-Stokes equations have been conducted on up to 24 processors
in~\cite{Trindade2004,Trindade2006}. While they successfully
established the general applicability of such a space-time parallel
approach for the Navier-Stokes equations, the obtained speedups were
ambiguous: Best speedups were achieved either with a pure
time-parallel or a pure space-parallel approach, depending on the
problem size.

In this paper we combine the Parareal-in-time and domain-decomposi\-tion-in-space
techniques and investigate this space-time parallel scheme by means
of solving a quasi-2D and a fully 3D driven cavity flow problem on a 
state-of-the-art HPC distributed memory architecture, using up to 2,048 cores. 
We demonstrate the capability of the approach to reduce time-to-solution below the saturation point of a
pure spatial parallelization. Furthermore, we show that
the addition of obstacles into the computational domain, leading to
more turbulent flow, leads to slower convergence of Parareal. This is
likely due to the reported stability issues for hyperbolic and convection-dominated problems,
see~\cite{FarhatEtAl2003,RuprechtKrause2012}.

\section{Physical model and its discretization and parallelization}
%subsection{Physical model}
The behavior of three-dimensional, incompressible Newtonian fluids is
described by the incompressible Navier-Stokes equations.
% defined on an
%open set $\Omega \subset \mathbb{R}^3$ with Lipschitz boundary
%$\Gamma:=\partial\Omega$ for the time interval $t\in (0,T],~T>0$.
In dimensionless form the according momentum- and continuum equation read
{\newline
\parbox{0.9\textwidth}{
\begin{align*}
	\partial_{t} \Tt{u} + \Tt{u} \cdot \nabla \Tt{u} 
	 &= \frac{1}{\textrm{Re}}\Delta \Tt{u} - \nabla p  \\
	\nabla \cdot \Tt{u} &= 0
\end{align*}}\hfill
\parbox{0.05\textwidth}{\begin{align} \label{eq:navierStokes}
\end{align}}\newline} with $\Tt{u} = (u,v,w)$ being the velocity field
consisting of the Cartesian velocity-components, $p$ being the
pressure and $\textrm{Re}$ the dimensionless Reynolds number.

%\subsection{Discretization}
The Navier-Stokes solver is based on the software-package
NaSt3DGP~\cite{NaSt3DGP,GriebelDornseiferNeunhoeffer:1998} and we
further extended it by an MPI-based implementation of
Parareal~\cite{LionsEtAl2001}. In NaSt3DGP, the unsteady
3D-Navier-Stokes equations are discretized via standard finite
volume/finite differences using the Chorin-Temam~\cite{Chorin,Temam}
projection method on a uniform Cartesian staggered mesh for robust
pressure and velocity coupling. A first order forward Euler scheme is
used for time discretization and as a building block for Parareal, see
the description in~\ref{subsubsec:Parareal}. Second order central
differences are used for the pressure gradient and diffusion. The
convective terms are discretized with a second order TVD
SMART~\cite{SMART} upwind scheme, which is basically a bounded
QUICK~\cite{QUICK} scheme. Furthermore, complex geometries are
approximated using a first order cell decomposition/enumeration
technique, on which we can impose slip as well as no-slip boundary
conditions. Finally, the Poisson equation for the pressure arising in
the projection step is solved using a BiCGStab~\cite{BiCGStab}
iterative method.

\subsection{Parallelization}
Both the spatial as well as the temporal parallelization are
implemented for distribu\-ted-memory machines using the MPI-library.
 The underlying algorithms are described in the following.

\subsubsection{Parallelization in space via domain decomposition}
We uniformly decompose the discrete computational domain $\Omega_{h}$
into $P$ subdomains by first computing all factorizations of $P$ into
three components, i.e. $P=P^x\cdot P^y \cdot P^z,$ with $P^x,P^y,P^z
\in \mathbb{N}$. 
%Then we minimize the following cost-function $C$ for
%communication with respect to all pre-computed factorizations of $P$,
%i.e. the space decomposition is generated in view of the overall
%surface area minimization of neighboring subdomains:
Then we use our pre-computed factorizations of $P$ as arguments
for the following cost function $C$ with respect to communication 
\begin{align} 
C(P^x,P^y,P^z) &= \frac{I}{P^x}\cdot\frac{J}{P^y} +
\frac{J}{P^y}\cdot\frac{K}{P^z} + \frac{I}{P^x}\cdot\frac{K}{P^z}
\end{align}
with $I, J$ and $K$ as the total number of grid-cells in $x$-, $y$-
and $z$-direction.
Finally, we apply that factorization for the domain decomposition for
which C is minimal, i.e. the space decomposition is generated in view
of the overall surface area minimization of neighboring subdomains.
%\begin{align}
%\textnormal{min } 
%C(P^x,P^y,P^z) &= \frac{I}{P^x}\cdot\frac{J}{P^y} +
%\frac{J}{P^y}\cdot\frac{K}{P^z} + \frac{I}{P^x}\cdot\frac{K}{P^z}\\ \nonumber 
%\\ 
%\textnormal{min}~C~\textnormal{  under side condition  }~~P &= P^x\cdot
%P^y \cdot P^z~~\textnormal{with}~~P^x,P^y,P^z \in \mathbb{N}~.
%\end{align}
%with $I, J$ and $K$ as the total number of grid-cells in $x$-, $y$-
%and $z$-direction. 
Here, $P$ is always identical to the number of processors $N_{\rm
p_{\rm space}}$, so that each processor handles one subdomain. Since
the stencil is five grid-points large for the convective terms and
three grid-points for the Poisson equation, each subdomain needs two
ghost-cell rows for the velocities and one ghost-cell row for the
pressure Poisson equation.  Thus our domain decomposition method needs
to communicate the velocities once at each time-step and the pressure
once at each pressure Poisson iteration.

\subsubsection{Parallelization in time with Parareal}\label{subsubsec:Parareal}
For a given time interval $[0,T]$, we introduce a coarse temporal mesh
\begin{equation}
	0 = t_{0} < t_{1} < \ldots < t_{N_{\rm c}} = T
\end{equation}
with a uniform time-step size $\Delta t = t_{i+1} - t_{i}$. Further,
we introduce a fine time-step $\delta t < \Delta t$ and denote by $N_{\rm
f}$ the total number of fine steps and by $N_{\rm c}$ the total number
of coarse steps, that is
\begin{equation}
	N_{\rm c} \Delta t = N_{\rm f} \delta t = T.
\end{equation}
Also assume that the coarse time-step is a multiple of the fine, so that
\begin{equation}
	\frac{\Delta t}{\delta t} =: N_{\rm r} \in \mathbb{N}.
\end{equation}
Parareal relies on the iterative use of two integration schemes: a
fine propagator $\mathcal{F}_{\delta t}$ that is computationally
expensive, and a coarse propagator $\mathcal{G}_{\Delta t}$ that is
computationally cheap. We sketch the algorithm only very briefly here, for
a more detailed description see for example~\cite{LionsEtAl2001}.

Denote by $\mathcal{F}(\Tt{y}, t_{n+1}, t_{n})$, $\mathcal{G}(\Tt{y},t_{n+1},t_{n})$
the result of integrating from an initial value $\Tt{y}$ at time
$t_{n}$ to a time $t_{n+1}$, using the fine or coarse scheme,
respectively. Then, the basic iteration of Parareal reads
\begin{equation}
	\label{eq:parareal}
	\Tt{y}^{k+1}_{n+1} = \mathcal{G}_{\Delta t}(\Tt{y}^{k+1}_{n},
	t_{n+1}, t_{n}) + \mathcal{F}_{\delta t}(\Tt{y}^{k}_{n},
	t_{n+1}, t_{n}) - \mathcal{G}_{\Delta t}(\Tt{y}^{k}_{n},
	t_{n+1}, t_{n})
\end{equation}
with super-scripts referring to the iteration index and $\Tt{y}_{n}$
corresponding to the approximation of the solution at time $t_{n}$.
Iteration \eqref{eq:parareal} converges to a solution
\begin{equation}
	\Tt{y}_{n+1} = \mathcal{F}_{\delta t}(\Tt{y}_{n}, t_{n+1}, t_{n}),
\end{equation}
that is a solution with the accuracy of the fine solver. Here, we
always perform some prescribed number of iterations $N_{\rm
it}$. We use a forward
Euler scheme for both $\mathcal{F}_{\delta t}$ and
$\mathcal{G}_{\Delta t}$ and simply use a larger time-step for the
coarse propagator. Experimenting with the combination of schemes of
different and/or higher order is left for future work.

Once the values $\Tt{y}^{k}_{n}$ in~\eqref{eq:parareal} from the
previous iteration are known, the computationally expensive
calculations of the values $\mathcal{F}_{\delta t}(\Tt{y}^{k}_{n},
t_{n+1}, t_{n})$ can be performed in parallel for multiple coarse
intervals $[t_{n},t_{n+1}]$.  In the pure time-parallel case, the
time-slices are distributed to $N_{\rm p_{\rm time}}$ cores assigned
for the time-parallelization. Note that in
the space-time parallel case, the time-slices are not handled by
single cores but by multiple cores, each handling one subdomain at
the specific time, see Section~\ref{subsubsec:spacetime}.

The theoretically obtainable speedup with Parareal is bounded by
\begin{equation}
	\label{eq:theoretSpeedup}
	s(N_{\rm p}) \leq \frac{N_{\rm p_{\rm time}}}{N_{\rm it}}
\end{equation}
with $N_{\rm p_{\rm time}}$ denoting the number of processors in the
temporal parallelization and $N_{\rm it}$ the number of iterations,
see for example \cite{Minion2010}. From \eqref{eq:theoretSpeedup} it
follows that the maximum achievable parallel efficiency  of the time parallelization is bounded by
$1/N_{\rm it}$. Parareal is hence considered as an additional
direction of parallelization to be used when the spatial
parallelization is saturated but a further reduction of
time-to-solution is required or desirable. Some progress has recently been made
deriving time-parallel schemes with less strict efficiency bounds~\cite{EmmettMinion2012,Minion2010}.

\subsubsection{Combined parallelization in space and time}\label{subsubsec:spacetime}
\begin{figure}[t]
  \centering
  \includegraphics[width=!,height=0.67\textwidth,angle=-90]{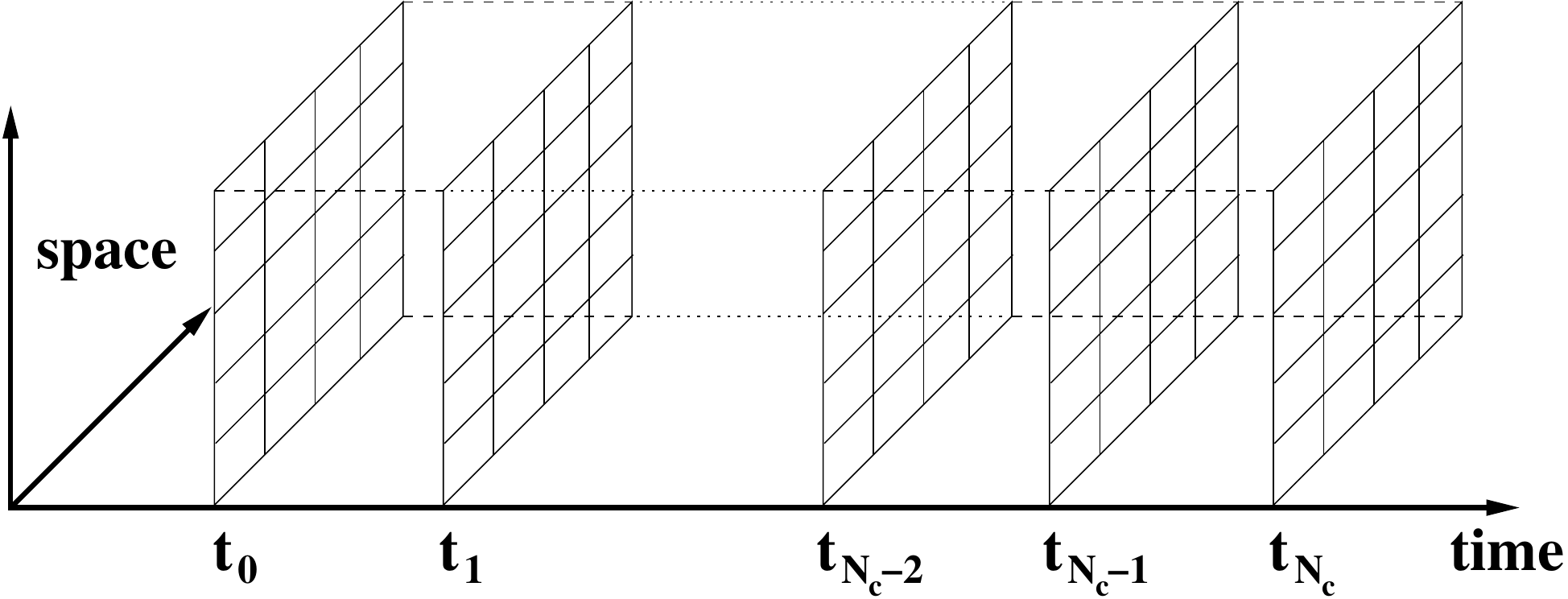}
  \caption{Decomposition of the time interval $[0,T]$ into $N_{\rm c}$
time-slices. The spatial mesh at each point $t_{i}$ is again
decomposed into $P$ subdomains, assigned to $N_{\rm p_{\rm space}}$
cores. Because the spatial parallelization does not need to
communicate across time-slices, the cores from every spatial mesh are
pooled into one MPI communicator.  Also, in the time parallelization,
only cores handling the same subdomain at different times have to
communicate.  Note that for readability the sketched spatial mesh is
2D, although the simulations use a fully 3D mesh.\label{fig:space-time}}
\end{figure}
In the combined space-time parallel approach as sketched in Figure
\ref{fig:space-time}, each coarse time interval
in~\eqref{eq:parareal} is assigned not to a single processor, but to
one MPI communicator containing $N_{\rm p_{\rm space}}$ cores, each
handling one subdomain of the corresponding time-slice. The total
number of cores is hence
\begin{equation}
  N_{\rm p_{\rm total}} = N_{\rm p_{\rm time}} \times N_{\rm p_{\rm space}}.
\end{equation}
Note that the communication in time in~\eqref{eq:parareal} is local in
the sense that each processor has only to communicate with the cores
handling the same subdomain in adjacent time-slices. Also, the spatial
parallelization is not communicating across time-slices, so that for
the evaluation of $\mathcal{F}$ or $\mathcal{G}$ in~\eqref{eq:parareal}, 
no communication between processors assigned to
different points in time is required. We thus organize all available
cores into two types of MPI communicators: (i) Spatial communicators
collect all cores belonging to the solution at one fixed time-slice,
but handling different subdomains. They correspond to the distributed
representation of the solution at one fixed time-slice. There are
$N_{\rm p_{\rm time}}$ spatial communicators and each contains $N_{\rm
p_{\rm space}}$ cores. \hbox{(ii) Time communicators} collect all cores dealing
with the same spatial subdomain, but at different time-slices. They
are used to perform the iterative update in~\eqref{eq:parareal} of the
local solution on a spatial subdomain. There are $N_{\rm p_{\rm
space}}$ time communicators, each pooling $N_{\rm p_{\rm time}}$
cores. No special attention was paid to how
 different MPI tasks are assigned to cores. 
Because of the very
different communication pattern of the space- and
time-parallelization, this can presumably have a significant effect on
the overall performance. More detailed investigation of the optimal
placement of tasks is planned for future studies with the here
presented code.

\section{Numerical examples}
\begin{figure}[t]
	\centering
	\includegraphics[width=0.327\textwidth, height=0.202\textheight]{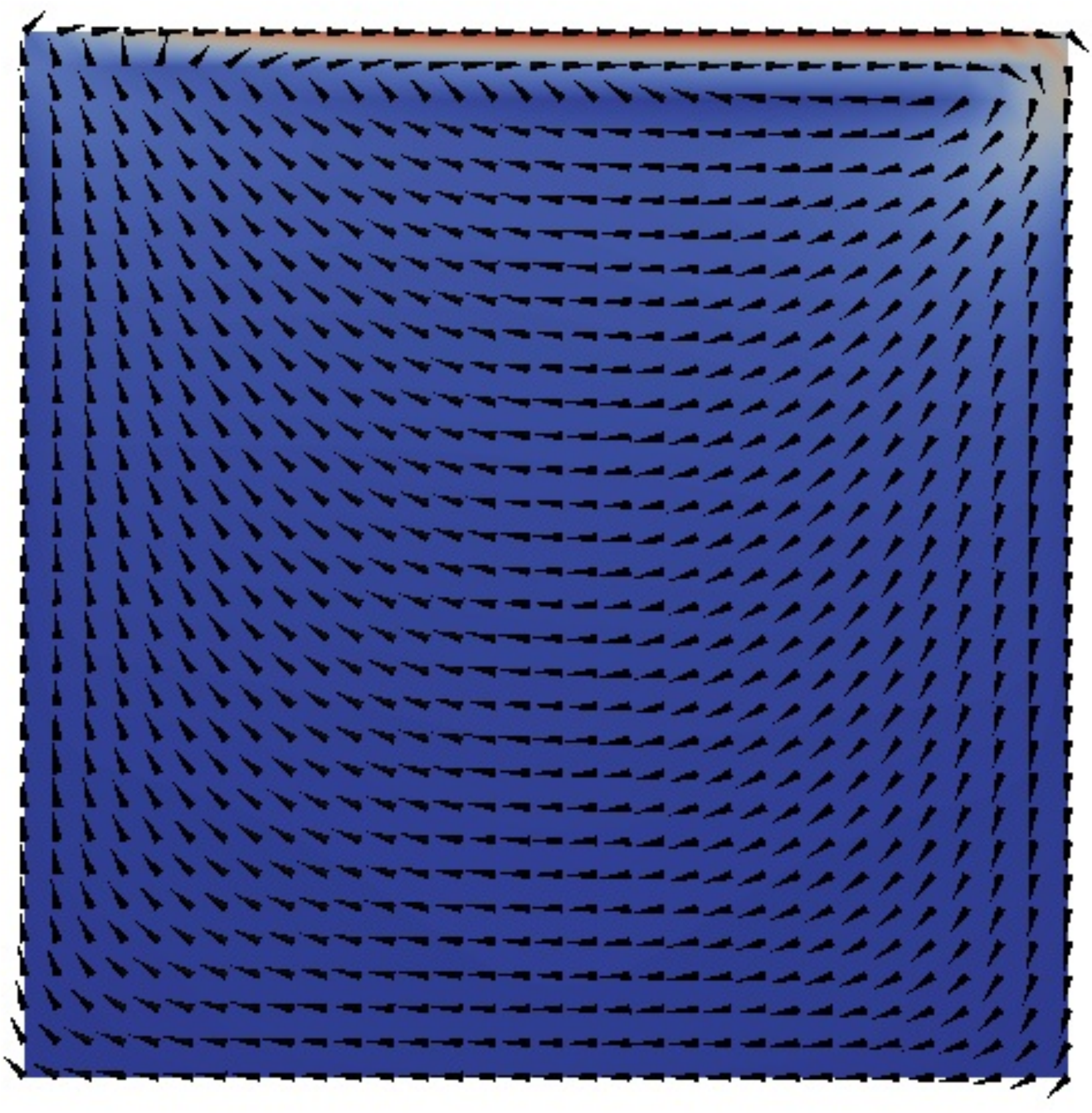}~
	\includegraphics[width=0.327\textwidth, height=0.202\textheight]{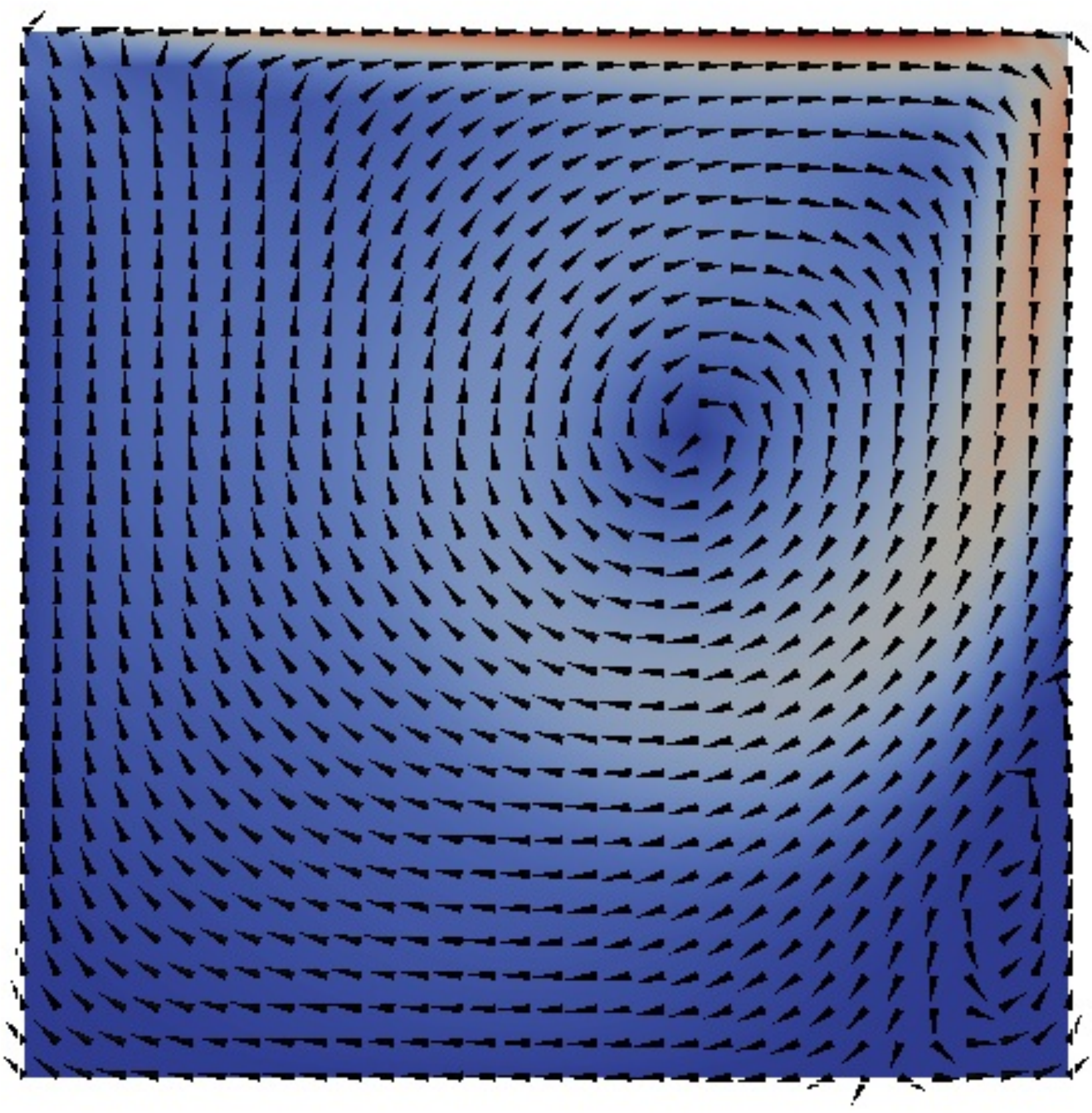}
	\includegraphics[width=0.327\textwidth, height=0.2015\textheight]{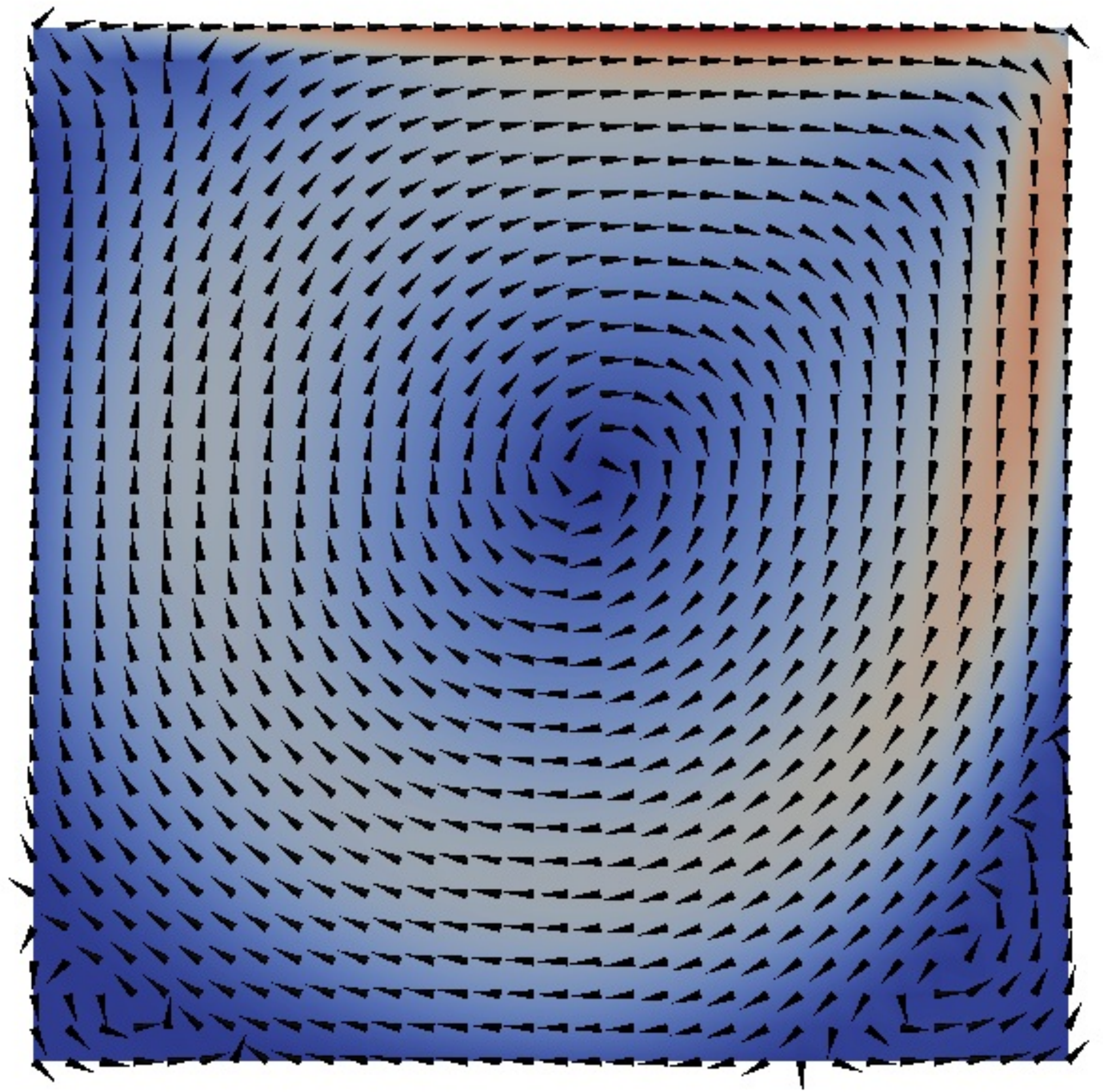}
	\caption{Simulation 1: Arrows and color plot (blue=0.0$-$red=0.75) of the
          Euclidean norm of the quasi two-dimensional driven cavity
          flow field along the center plane at three points in time
          $t=0.8$ and $t=8.0$ and $t=T=80.0$.\label{fig:drivenCavity}}
\end{figure}
\begin{figure}[t]
	\centering
	\includegraphics[width=0.38\textwidth]{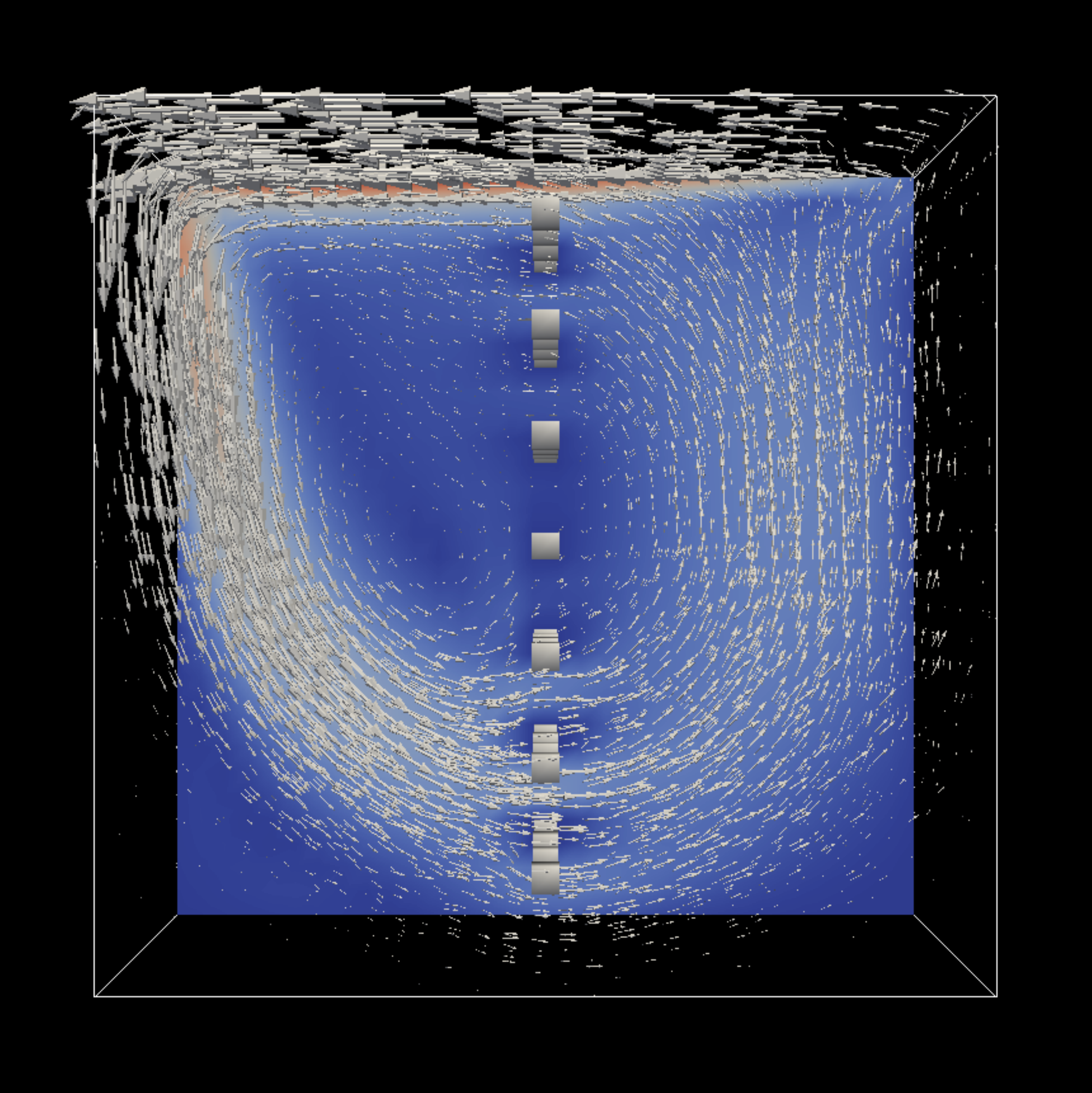}~~~~~~~~
	\includegraphics[width=0.38\textwidth]{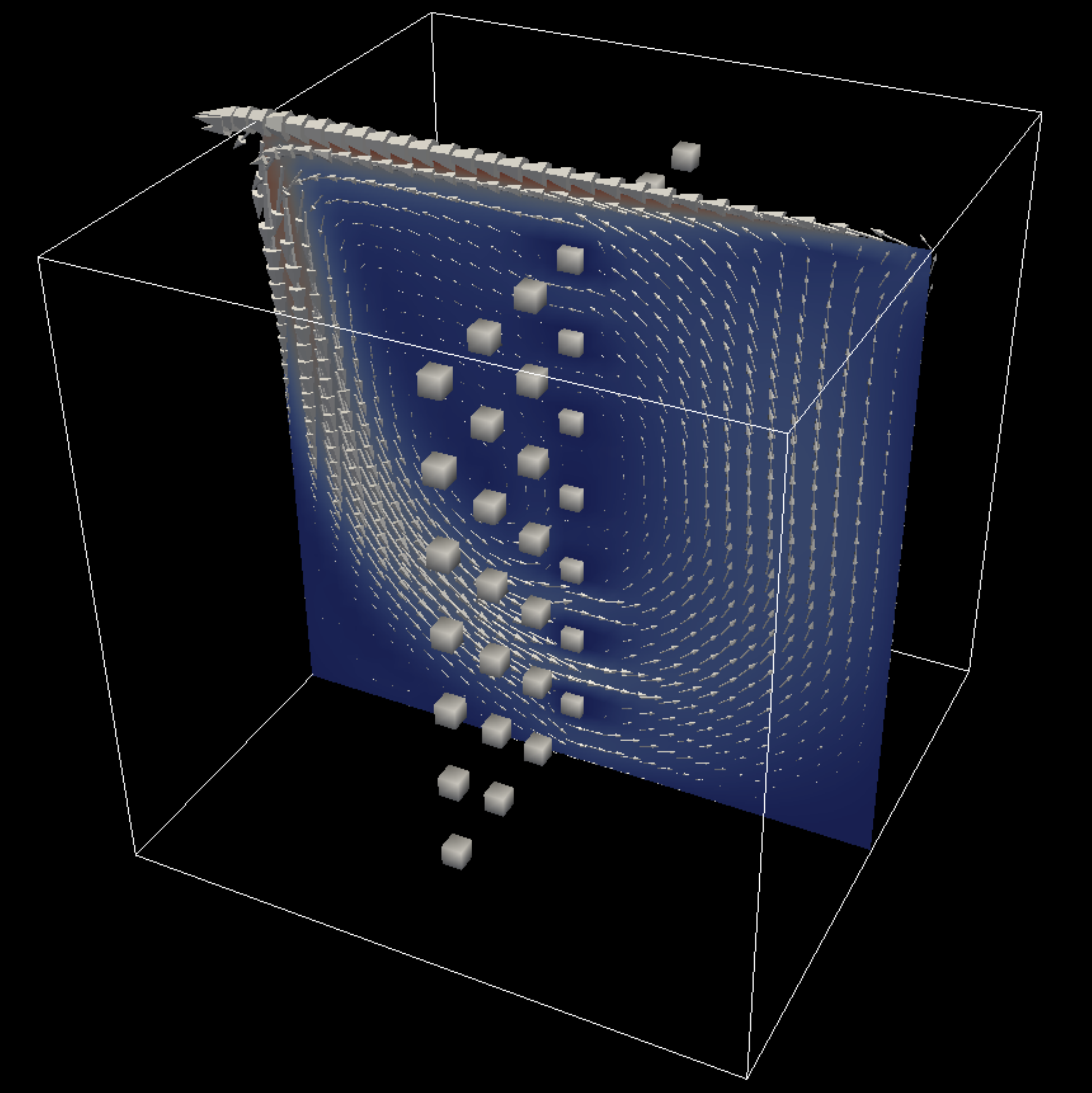}
	\caption{Simulation 2: Arrows and color plot (blue=0.0$-$red=0.75) of the
          Euclidean norm of the fully three-dimensional driven cavity
          flow field with obstacles along the center plane at
          $t=T=24.0$.\label{fig:drivenCavityObs}}
\end{figure}
In the following, we investigate the performance of the space-time
parallel approach for two numerical examples. The first is the
classical driven-cavity problem in a quasi-2D
setup. Figure~\ref{fig:drivenCavity} shows the flow in a $xy$-plane
with $z = 0.05$ at times $t=0.8,~t=8.0$ and $t=T=80.0$.  The second
example is an extension where $49$ obstacles (cubes) are inserted into
the domain along the median plane, leading to fully 3D
flow. Figure~\ref{fig:drivenCavityObs} sketches the obstacles and the
flow at time $t=T=24$. Both problems are posed on a 3D-domain with
periodic boundary conditions in $z$-direction (note that $x$ and $z$
are the horizontal coordinates, while $y$ is the vertical
coordinate). Initially, velocity and pressure are set to zero.  At the
upper boundary, a tangential velocity $u_{\rm boundary}=1$ is
prescribed, which generates a flow inside the domain as time
progresses.  No-slip boundary conditions are used at the bottom and in
the two $yz$-boundary planes located at $x=0$ and $x=1$ as well as on
the obstacles. The parameters for the two simulations are summarized
in Table~\ref{tab:simPar}.  To assess the temporal discretization
error of $\mathcal{F}_{\delta t}$, the solution is compared to a
reference solution computed with $\mathcal{F}_{\delta t/10}$, giving a
maximum error of $1.2 \times 10^{-5}$ for the full 3D flow with
obstacles. That means that once the iteration of Parareal has reduced
the maximum defect between the serial and parallel solution below this
threshold, the time-parallel and time-serial solution are of
comparable accuracy. We use this threshold also for the quasi-2D
example, bearing in mind that the simpler structure of the flow in
this case most likely renders the estimate too conservative. The code
is run on a Cray XE6 at the Swiss National Supercomputing Centre,
featuring a total of 1496 nodes, each with two 16-core 2.1GHz AMD
Interlagos CPUs and 32GB memory per node. Nodes are connected by a
Gemini 3D torus interconnect and the theoretical peak performance is
402 TFlops.
\begin{table}[t]
  \caption{Simulation parameters for the quasi-2D driven cavity flow (Simulation 1) and the fully 3D driven cavity flow
with obstacles (Simulation 2).\label{tab:simPar}}
\centering
 \begin{tabular}{c@{ \ : } c@{ \ = \ } c | c@{ \ : } c@{ \ = \ } c}
   Sim. 1 & $\Omega_{h}$ & $[0,1]\times [0,1] \times [0,0.1]$ 
   &~Sim. 2 & $\Omega_{h}$ & $[0,1] \times [0,1] \times [0,1]$\\
   Sim. 1 & $N_{x} \times N_{y} \times N_{z}$ & $32 \times 32 \times 5$
   &~Sim. 2 & $N_{x} \times N_{y} \times N_{z}$ & $32 \times 32 \times 32$  \\
   Sim. 1 & $T$ & $80$ 
   &~Sim. 2 & $T$ & $24$\\ 
   Both & $\Delta t$ & $0.01$ 
   &~Both & $\delta t$ & $0.001$\\
   Both & $\textrm{Re}$ & $1000$ 
   &~Both & $u_{\rm boundary}$ & $1$\\
   Sim. 1 & $N_{\rm p_{\rm space}}$ & $1,2,4,8$ 
   &~Sim. 2 & $N_{\rm p_{\rm space}}$ & $1,\ldots,128$\\
   Sim. 1 & $N_{\rm p_{\rm time}}$ & $4,8,16$
   &~Sim. 2 & $N_{\rm p_{\rm time}}$ & $8,16,32$
   \end{tabular}
\end{table}
\begin{figure}[b]
	\centering
	\includegraphics[width=0.325\textwidth]{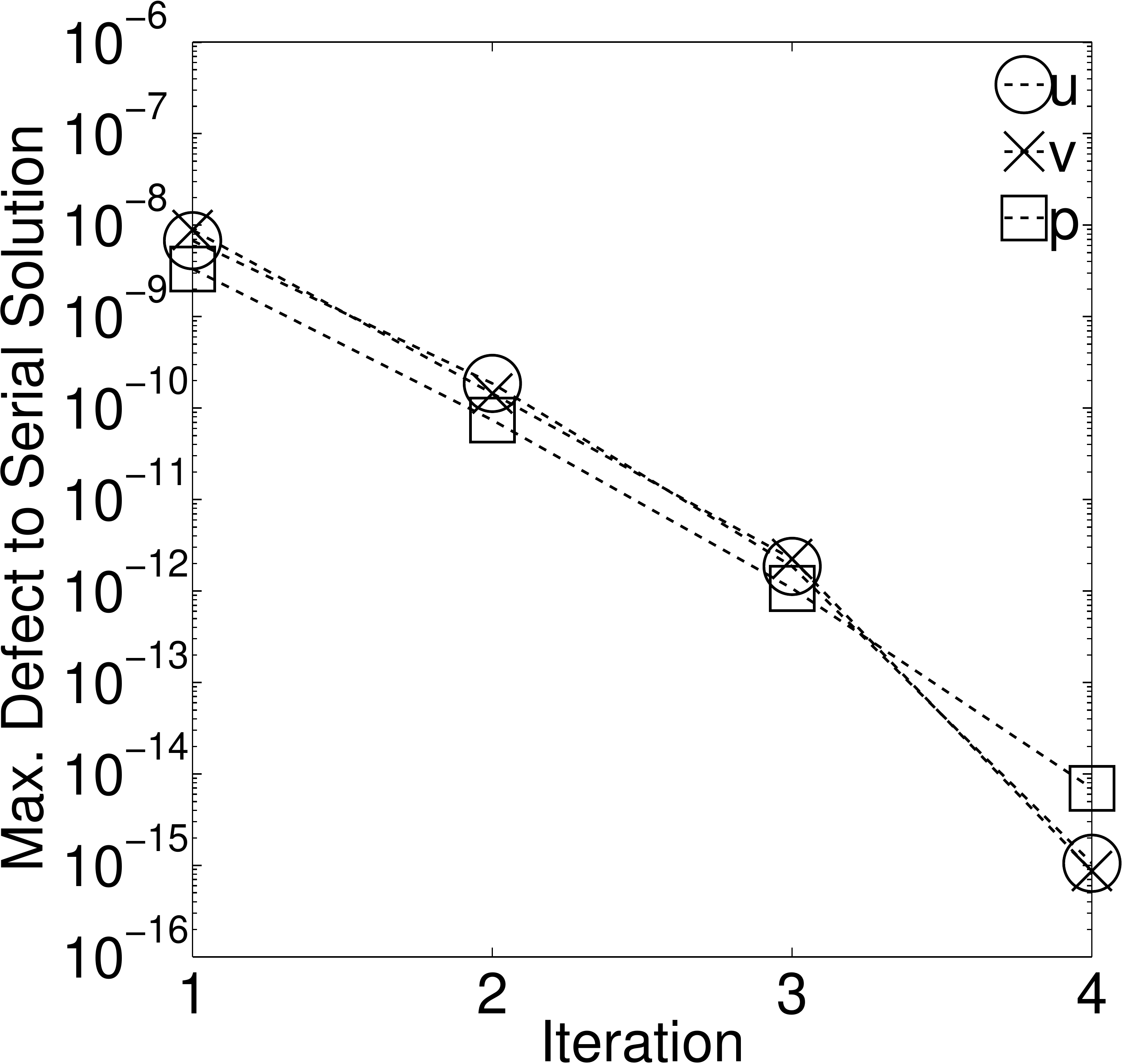}
	\includegraphics[width=0.325\textwidth]{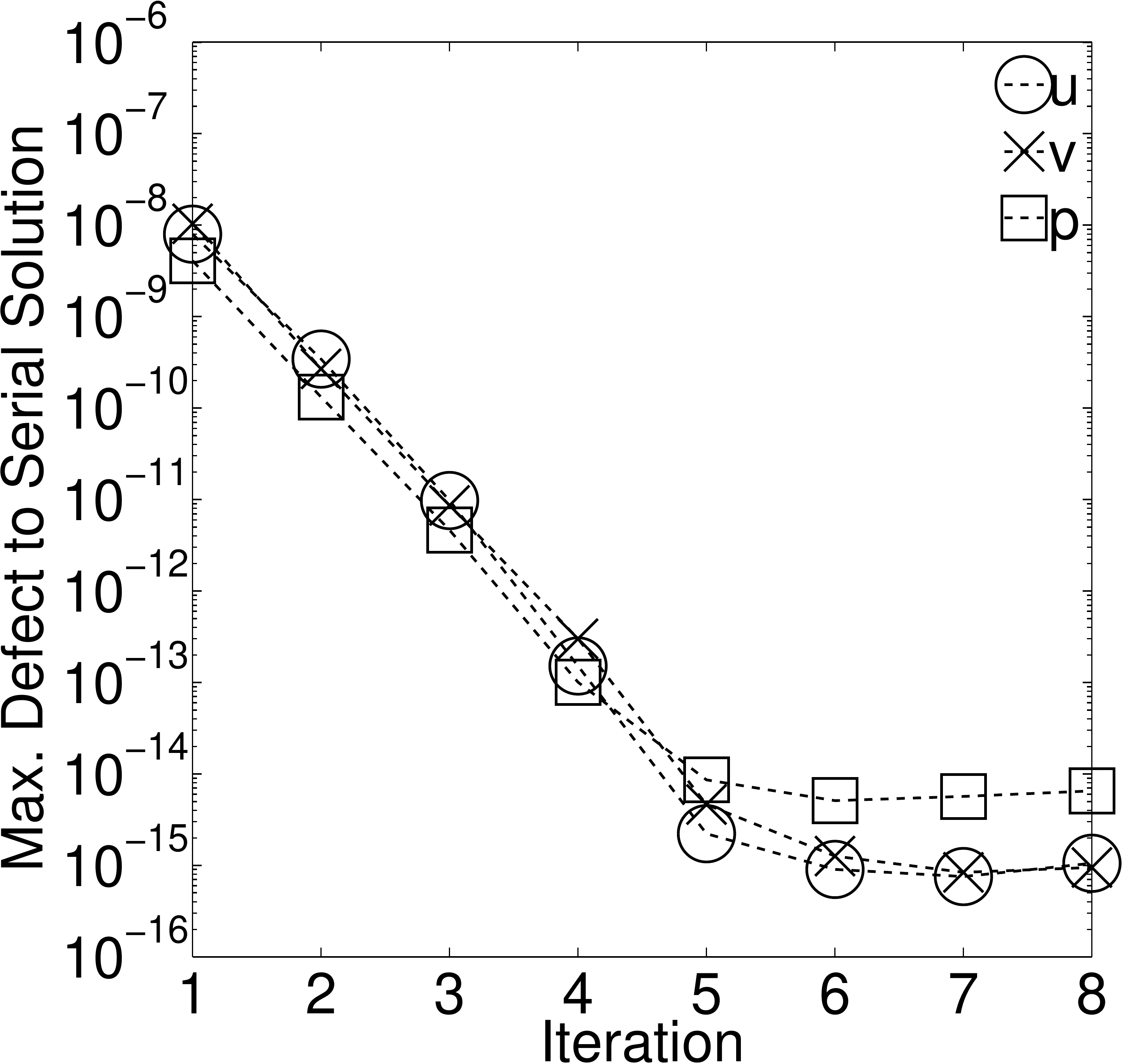}
	\includegraphics[width=0.325\textwidth]{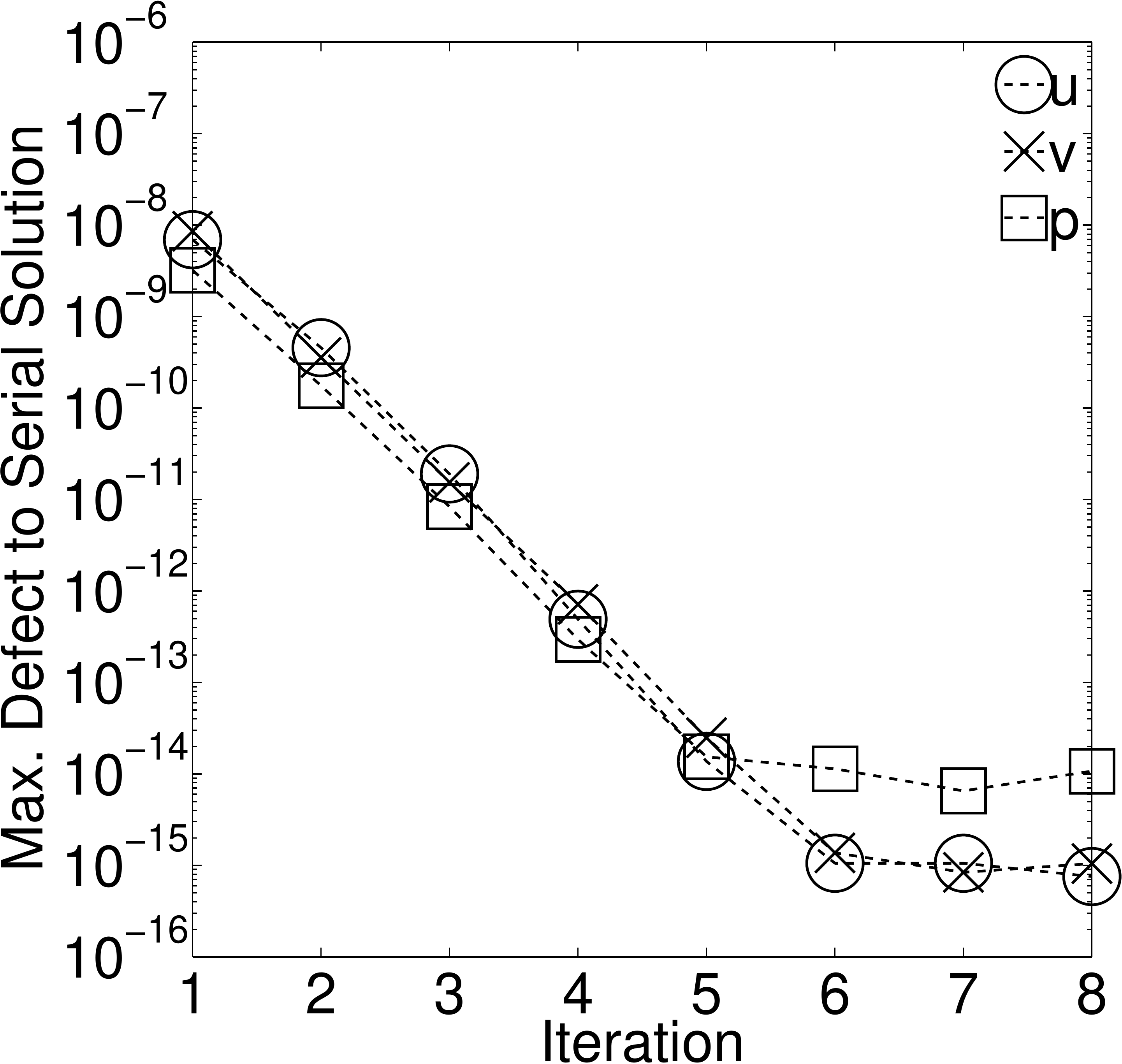}
	\caption{Maximum difference to time-serial solution versus
          number of Parareal iterations for the Cartesian velocity
          $(u,v)$ and pressure $p$ of the quasi 2D driven cavity
          problem at time $t = 80.0$ for $N_{p_{\rm time}}=4$ (left),
          $N_{p_{\rm time}}=8$ (middle) and $N_{p_{\rm time}}=16$
          (right).\label{fig:drivenCavityErrors}}
\end{figure}
\subsection{Quasi-2D driven cavity flow}
\begin{figure}[t]
	\centering
	\includegraphics[width=0.495\textwidth]{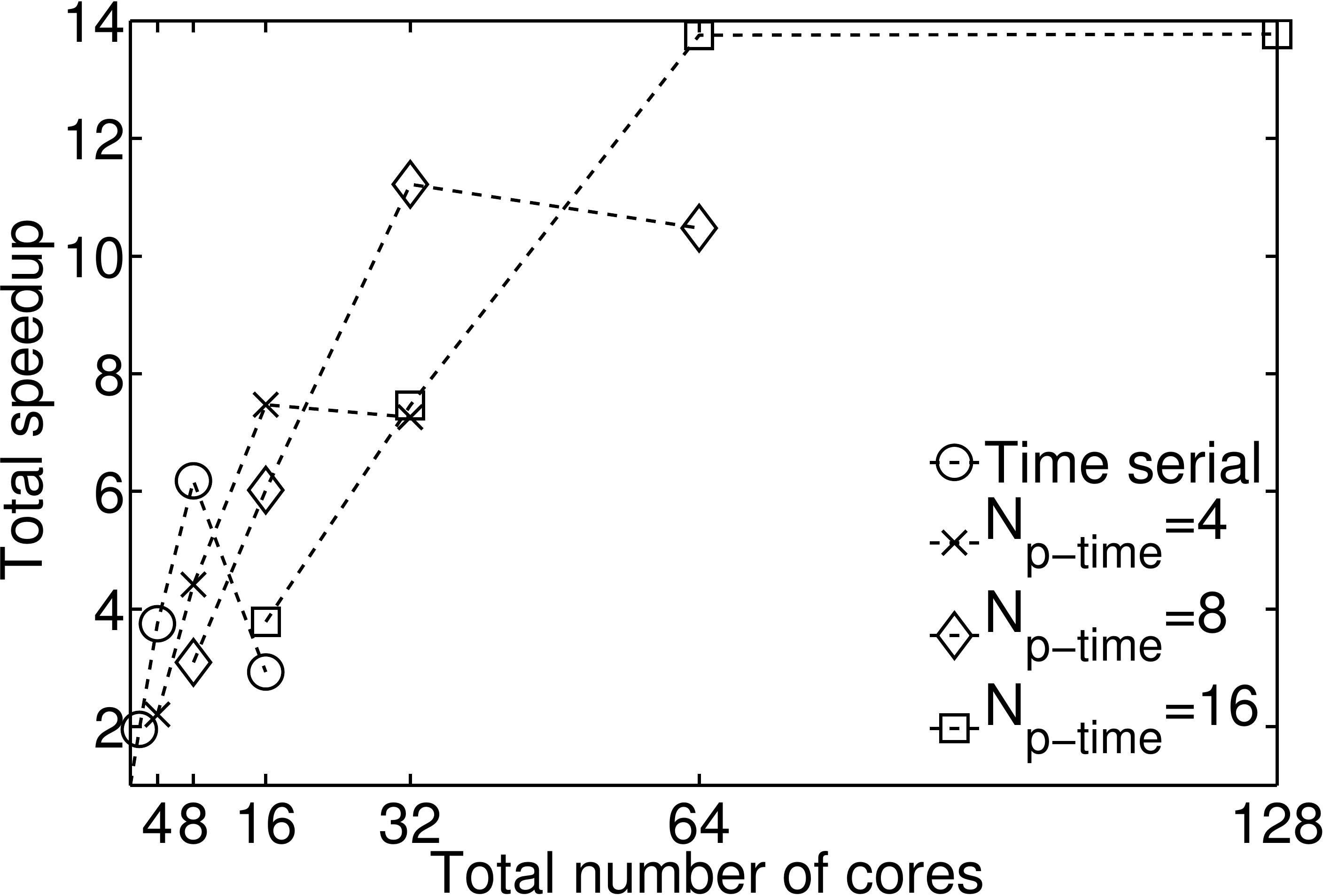}
	\includegraphics[width=0.495\textwidth]{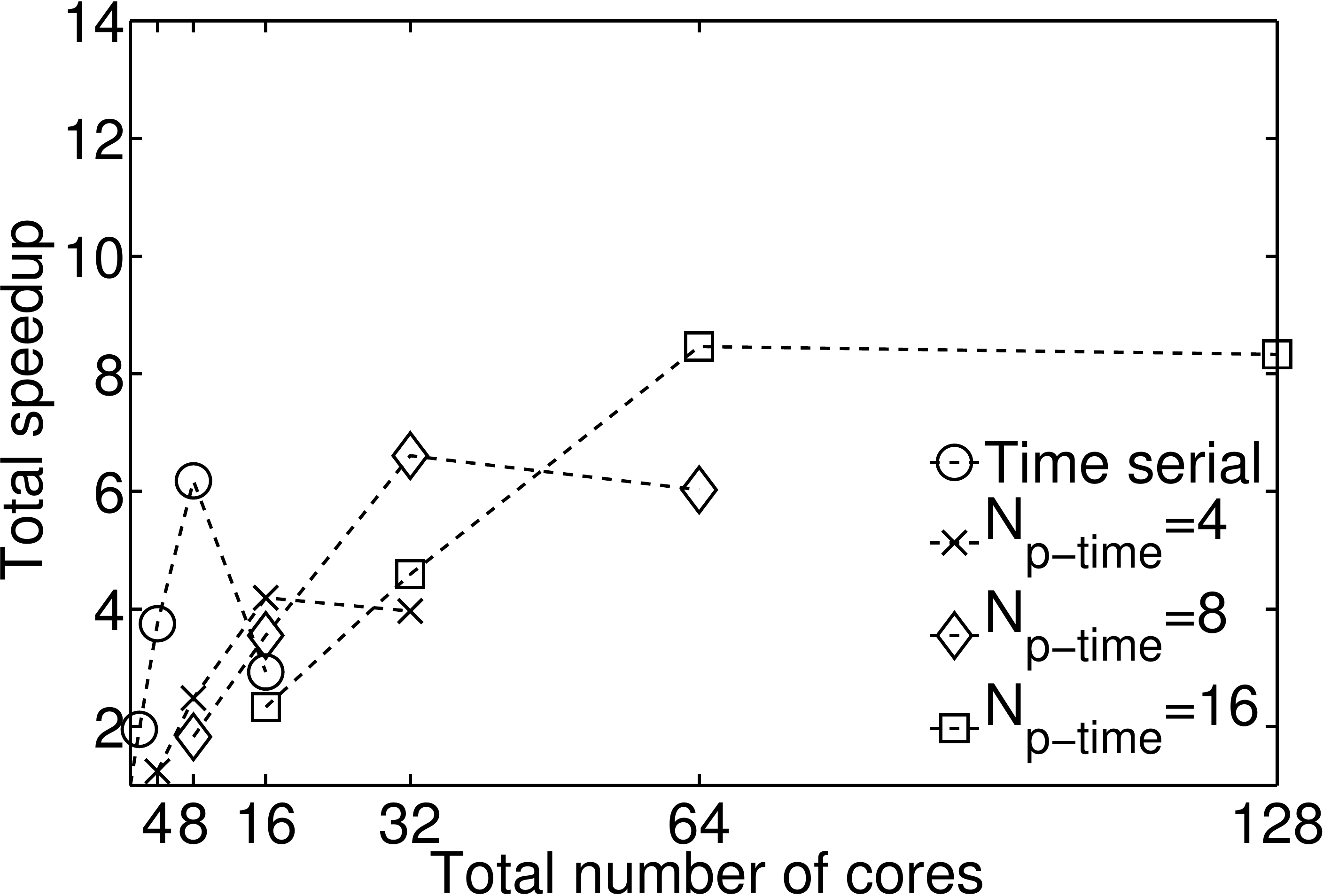}
	\caption{Total speedup of the combined space-time parallelization for quasi 2D driven-cavity flow with $N_{\rm it}=1$ (left) and
$N_{\rm it}=2$ (right) iterations.\label{fig:drivenCavitySpeedups}}
\end{figure}

Figure~\ref{fig:drivenCavityErrors} shows the maximum difference between the
time-parallel and the time-serial solution at the end of the simulation versus the number of iterations
of Parareal. In all three cases, the error decreases exponentially with $N_{\rm it}$. The threshold of $1.2 \times 10^{-5}$ is reached
after a single iteration, indicating that the performance of Parareal could probably be optimized by using a larger $\Delta t$.
Figure~\ref{fig:drivenCavitySpeedups} shows the total speedup provided by the time-serial scheme
running $\mathcal{F}_{\delta t}$ with only space-parallelism (circles) as well as by the
space-time parallel method for different values of $N_{p_{\rm time}}$. All speedups are measured against
the runtime of the time-serial solution run on a single core. The pure spatial parallelization reaches a maximum speedup of a little over 6 using 8 cores. For $N_{\rm it}=1$,
the space-time parallel scheme reaches a speedup of 14 using 64 cores. This amounts to a speedup of roughly $14 / 6 \approx 2.33$ from Parareal alone.
For $N_{\rm it}=2$ the speedup  is down to 8, but still noticeably larger than the saturation point of the pure space-parallel method. Note that because of the limited efficiency of the time parallelization, the slopes of the space-time parallel scheme are lower for larger values of $N_{p_{\rm time}}$.

\subsection{Full 3D driven cavity flow with obstacles}
\begin{figure}[t]
	\centering
	\includegraphics[width=0.32\textwidth]{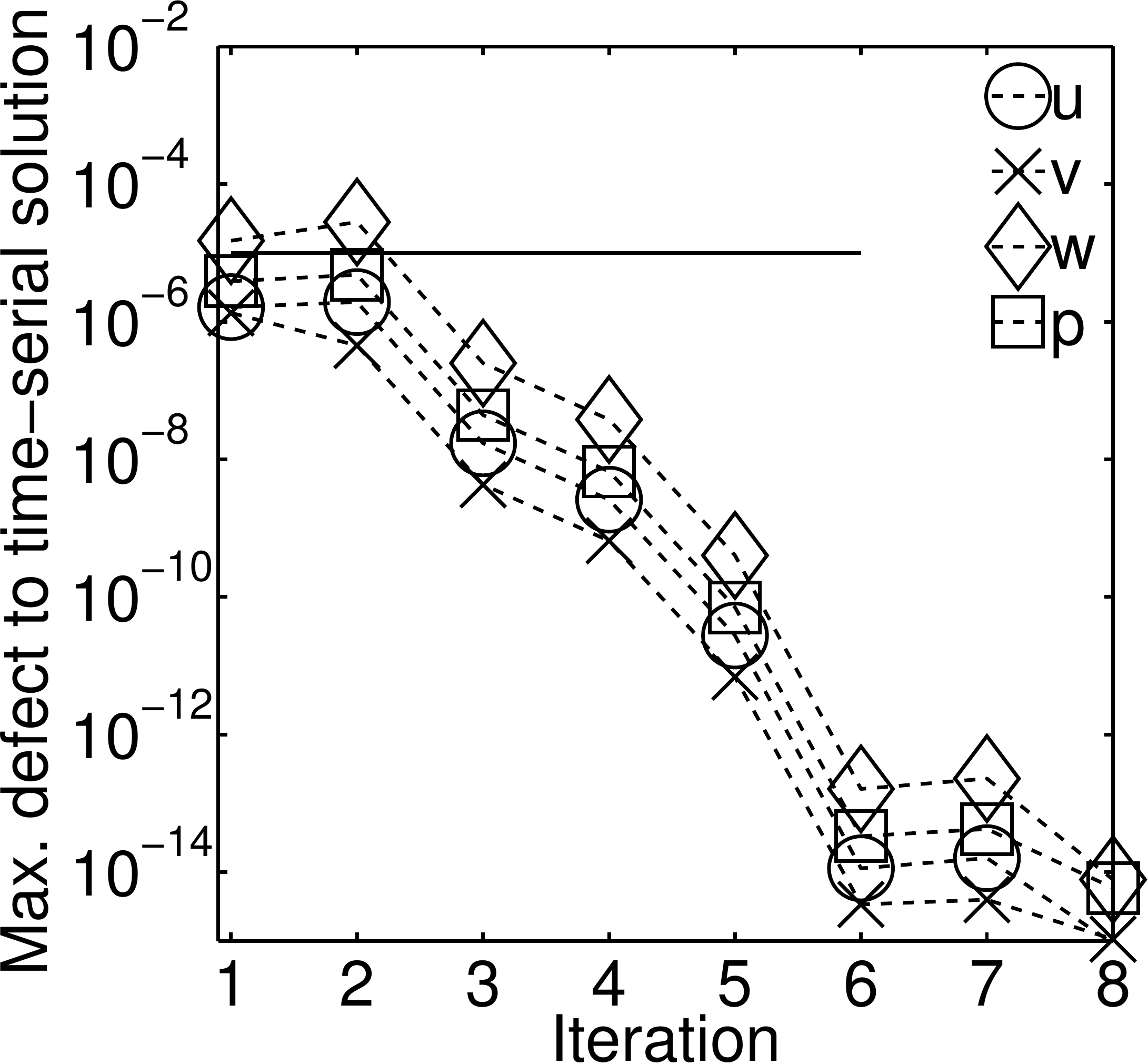}
	\includegraphics[width=0.32\textwidth]{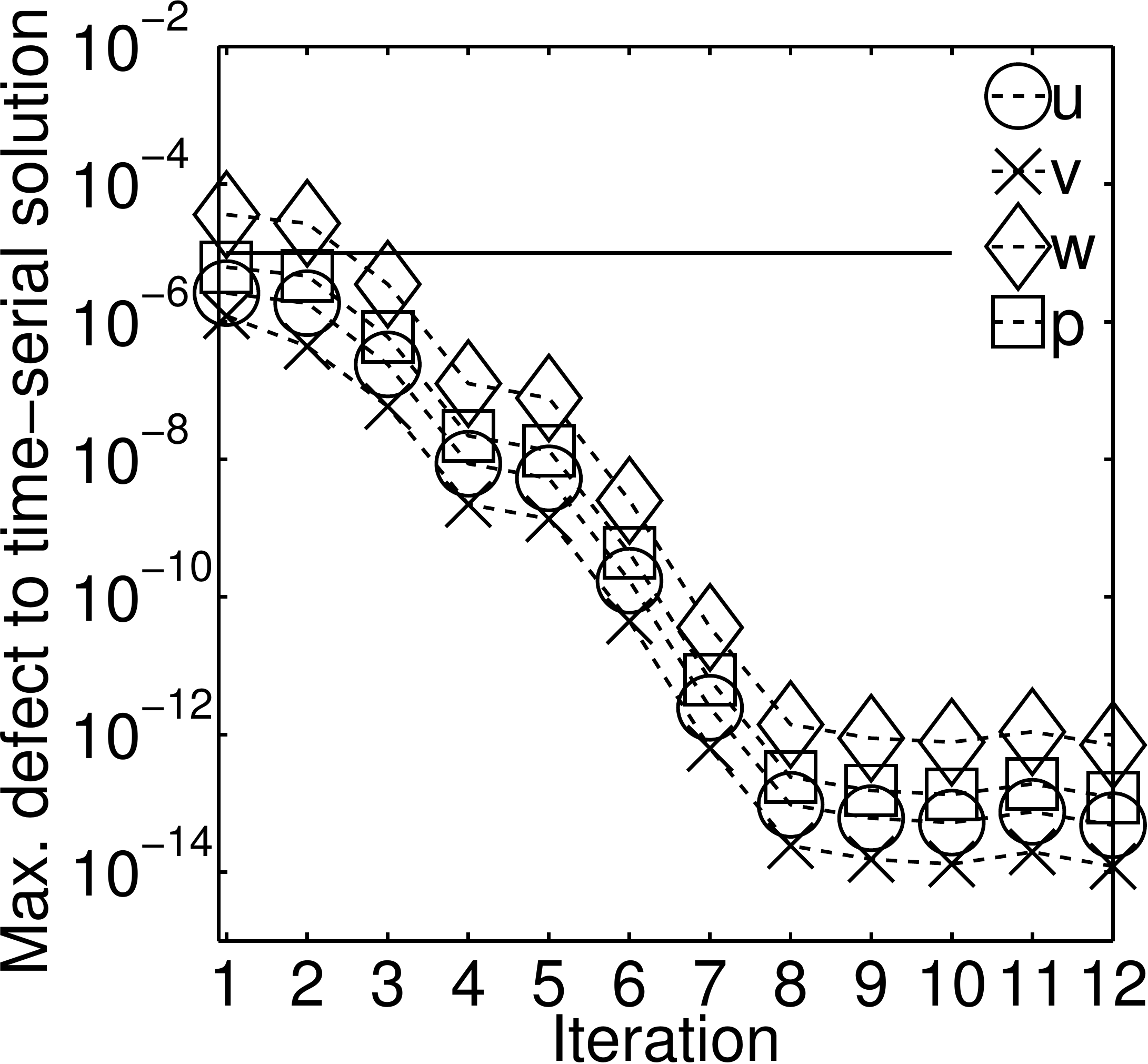}
	\includegraphics[width=0.32\textwidth]{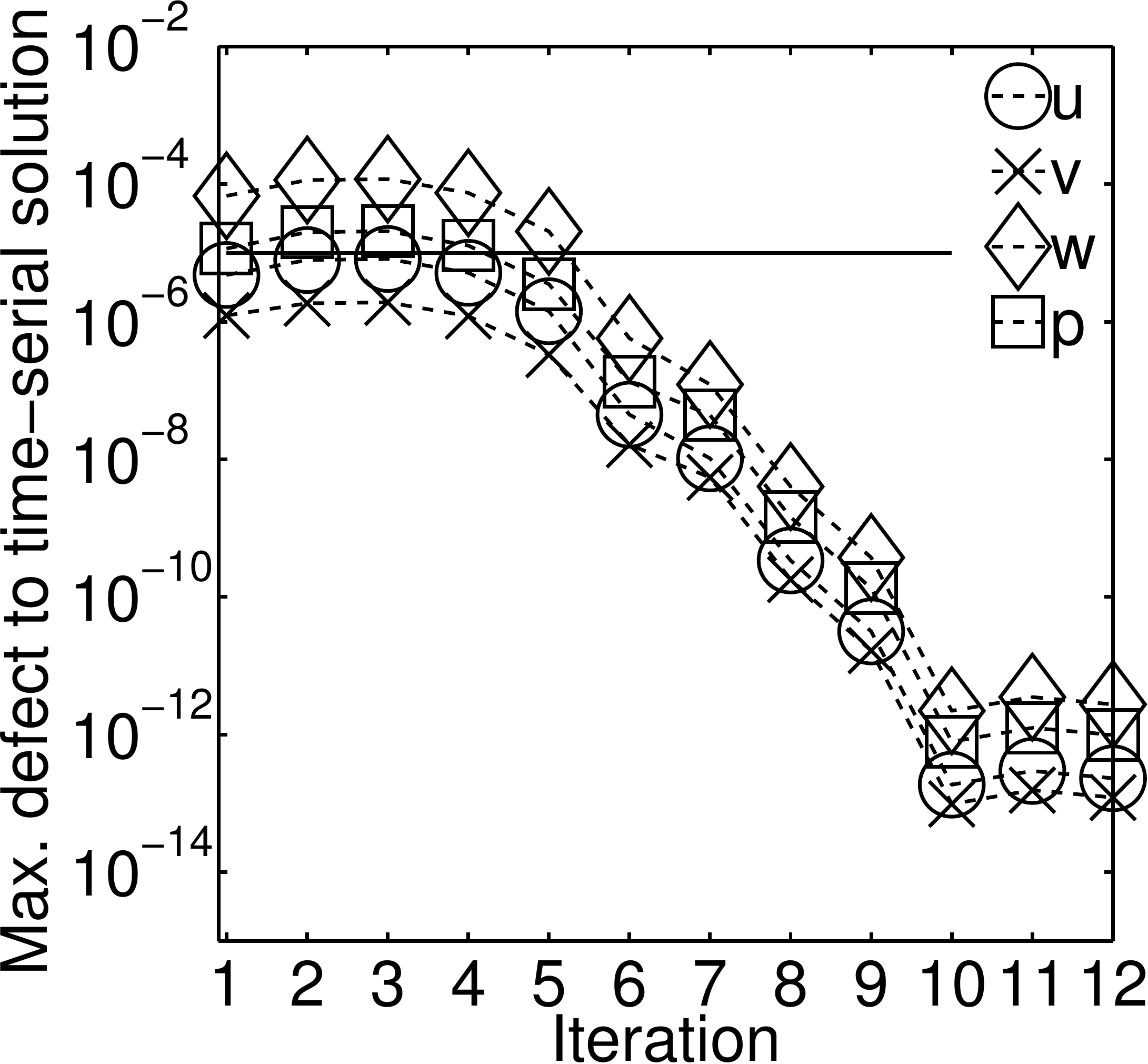}
	\caption{Maximum difference to time-serial solution at end of simulation versus number of Parareal iterations 
for the 3D driven cavity flow with obstacles for $N_{p_{\rm time}}=8$ (left), $N_{p_{\rm time}}=16$ (middle), $N_{p_{\rm time}}=32$ (right). The horizontal line
indicates an error level of $10^{-5}$.\label{fig:3Derror}}
\end{figure}

\begin{figure}[t]
	\centering
	\includegraphics[width=0.495\textwidth,height=0.32\textwidth]{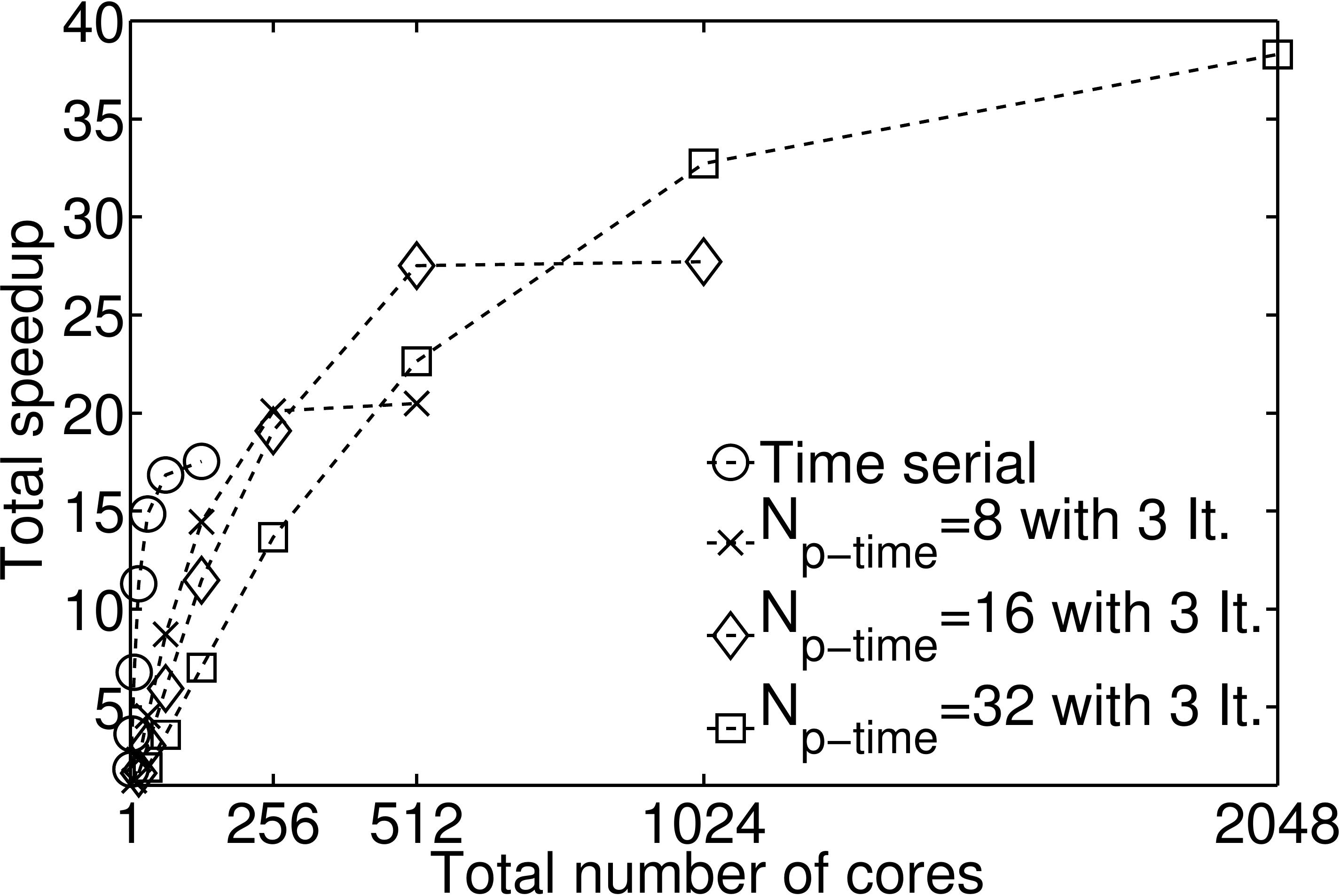}
	\includegraphics[width=0.495\textwidth,height=0.32\textwidth]{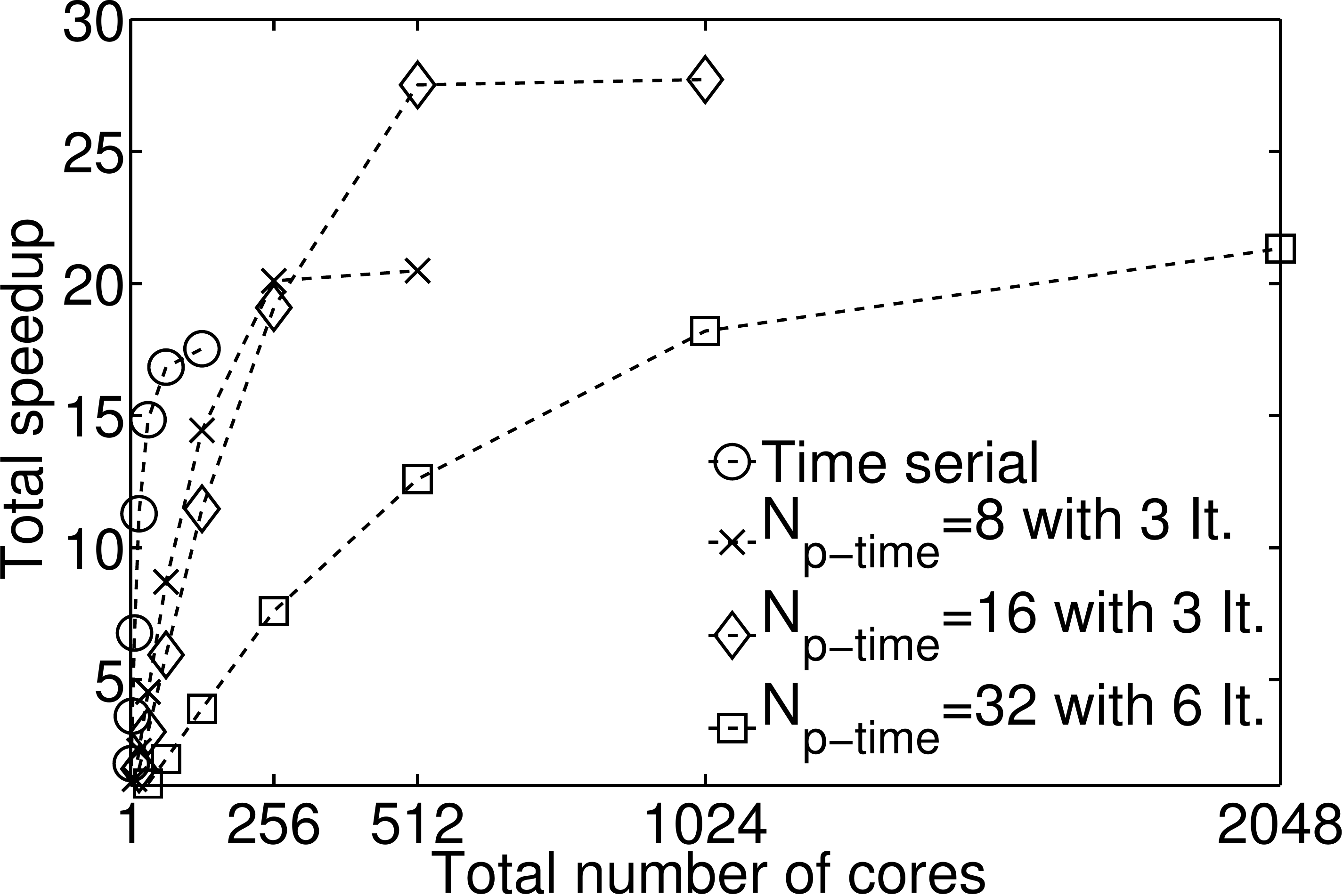}
	\caption{Total speedup of the combined space-time parallelization for 3D cavity flow with obstacles for a fixed number of $N_{\rm it}=3$ iterations (left) and  a number of iterations chosen to achieve a defect below $10^{-5}$ in all solution components (right). Note that the solutions in the left figure are not comparable in accuracy. \label{fig:speedup}}
\end{figure}

Depending on the number of Parareal iterations for three different
values of $N_{p_{\rm time}}$ Figure \ref{fig:3Derror} shows the
maximum difference between the time-parallel and the time-serial
solution in terms of the 3D-Cartesian velocity $(u,v,w)$ and pressure
$p$. In general, as in the quasi-2D case, the error decays
exponentially with the number of iterations, but now, particularly
pronounced for $N_{p_{\rm time}}=32$, a small number of iterations has
to be performed without large effect before the error starts to
decrease. This is likely due to the increased turbulence caused by the
obstacles, as it is known that Parareal exhibits instabilities for
advection dominated problems or hyperbolic
problems~\cite{Gander2007,RuprechtKrause2012}. A more detailed
analysis of the performance of Parareal for turbulent flow and larger
Reynolds numbers is left for future work. Figure~\ref{fig:speedup}
shows the total speedup measured against the runtime of the solution
running $\mathcal{F}_{\delta t}$ serially with $N_{\rm p_{\rm
    space}}=1$. The time-serial-line (circles) shows the speedup for a
pure spatial parallelization, which scales to $N_{\rm p_{\rm space}} =
16$ cores and then saturates at a speedup of about 18.  Adding
time-parallelism can significantly increase the total speedup, to
about 20 for $N_{p_{\rm time}}=4$, about 27 for $N_{p_{\rm time}}=8$
and to almost 40 for $N_{p_{\rm time}}=16$ for a fixed number of
$N_{\rm it}=3$ iterations (left figure). However, as can be seen from
Figure~\ref{fig:3Derror}, the solution with $N_{\rm p} = 32$ is
significantly less accurate. The right figure shows the total speedup
for a number of iterations adjusted so that the defect of Parareal in
all cases is below $10^{-5}$ in all solution components
(cf. Figure~\ref{fig:3Derror}). This illustrates that there is a
sweet-spot in the number of concurrently treated time-slices: At some
point the potential increase in speedup is offset by the additional
iterations required. In the presented example, the solution with
$N_{\rm p}=16$ is clearly more efficient than the one with $N_{\rm
  p}=32$.


\section{Conclusions}
A space-time parallel method, coupling Parareal with spatial domain 
decomposition, is presented and used to solve the three-dimensional,
time-dependent, incompressible Navier-Stokes equations.
%Both the space and the time parallel scheme are MPI-based and a
%description of the organization of the different processes into
%communicators is given.
Two setups are analyzed: A quasi-2D driven cavity example and an
extended setup, where obstacles inside the domain lead to a fully 3D
driven flow. The convergence of Parareal is investigated and speedups of the
space-time parallel approach are compared to speedups from a pure
space-parallel scheme. It is found that Parareal converges very
rapidly for the quasi-2D case. It also converges in the 3D case,
although for larger numbers of Parareal time-slices, convergence
starts to stagnate for the first few iterations, likely because of the
known stability issues of Parareal for advection dominated
flows. Results are reported from runs on up to 128 nodes with a total
of 2,048 cores on a Cray XE6, illustrating the feasibility of the
approach for state-of-the-art HPC systems. The results clearly
demonstrate the potential of time-parallelism as an additional
direction of parallelization to provide additional speedup after a
pure spatial parallelization reaches saturation. While the limited
parallel efficiency of Parareal in its current form is a drawback, we
expect the scalability properties of Parareal to direct future
research towards modified schemes with relaxed efficiency bounds.

\begin{acknowledgement}
This research is funded by the Swiss "High Performance and High
Productivity Computing" initiative HP2C. Computational resources were
provided by the Swiss National Supercomputing Centre CSCS.
\end{acknowledgement}
%
%\section*{Appendix}
%\addcontentsline{toc}{section}{Appendix}

\bibliographystyle{spmpsci}
\bibliography{Parareal,Misc}

\end{document}